\documentclass{PoS}
\usepackage{enumitem}
\setlist{nolistsep}

\title{Monte Carlo Performance Studies of Candidate Sites for the Cherenkov Telescope Array}

\ShortTitle{MC Performance Studies of Candidate Sites for CTA}

\author{\speaker{G. Maier}$^{a}$\, L. Arrabito$^{b}$, K. Bernl\"ohr$^{c}$,  J.Bregeon$^{b}$, F. Di Pierro$^{d}$, T. Hassan$^{e}$, T. Jogler$^{f}$, J. Hinton$^{c}$, A.  Moralejo$^{g}$, M.Wood$^{f}$
        for the CTA Consortium\footnote{Full consortium author list at http://cta-observatory.org}\\
        E-mail: \email{gernot.maier@desy.de}

{\footnotesize
$^{a}$DESY, Platanenallee 6, D-15738 Zeuthen, Germany; 
$^{b}$Laboratoire Univers et Particules de Montpellier - UMR5299, Universit\'e de Montpellier - CNRS/IN2P3, Place Eug\`ene Bataillon - CC 72, 34095 Montpellier C\'edex 05 France;
$^{c}$Max-Planck-Institut f\"ur Kernphysik, Heidelberg, Germany;
$^{d}$INAF - Osservatorio Astrofisico di Torino, Via P. Giuria 1, 10125 Torino, Italy;
$^{e}$Dpto. de F\'isica At\'omica, Molecular y Nuclear, Universidad Complutense de Madrid, Spain;
$^{f}$SLAC, USA;
$^{g}$IFAE, campus UAB, E-08193 Bellaterra, Spain
}}

\abstract{The Cherenkov Telescope Array (CTA) is the next-generation gamma-ray observatory with sensitivity in the energy range from 20 GeV to beyond 300 TeV. 
CTA is proposed to consist of two arrays of 40-100 imaging atmospheric Cherenkov telescopes, with one site located in each of the Northern and Southern Hemispheres. 
The evaluation process for the candidate sites for CTA is supported by detailed Monte Carlo simulations, which take different attributes like site altitude and geomagnetic field configuration into account. 
In this contribution we present the comparison of the sensitivity and performance of the different CTA site candidates for the measurement of very-high energy gamma rays.}

\FullConference{The 34th International Cosmic Ray Conference,\\
		30 July- 6 August, 2015\\
		The Hague, The Netherlands}

\begin{document}


\section{Introduction}

The Cherenkov Telescope Array (CTA) is the future gamma-ray observatory aiming at an order of magnitude improvement in flux sensitivity compared to current arrays of imaging atmospheric Cherenkov telescopes.
CTA will give deep and unprecedented insight into the non-thermal high-energy Universe.
The utilisation of  three different telescope sizes will provide a very broad energy range from as low as 20 GeV to beyond 300 TeV. 
The observatory is proposed to consist of two arrays, with one site located in each of the Northern and Southern Hemispheres. 
The baseline array for {\em CTA South} consists of 4 large-sized telescopes (LSTs), 25 mid-sized telescopes (MSTs) and about 70 small-sized telescopes (SSTs).
The northern array ({\em CTA North}) is somewhat smaller with 4 LSTs and 15 MSTs.
The site choices have significant impact on the overall science performance of the observatory.
The suitability of a site is given by 
\vspace{2mm}
\begin{itemize}
\item the  average annual observing time, determined by the number of night hours without moon or partial moon and the fraction of nights with suitable atmospheric conditions (typically between 1000 and 1300 hours / year);
\item the science performance (presented in this contribution), which is derived from site attributes like altitude, geomagnetic field, night-sky background level and  aerosol optical depth;
\item the risks, construction and operating costs.
\end{itemize}

\section{Monte Carlo Simulations of Site Candidates for CTA}

The CTA site evaluation studies on the science performance presented in the following uses detailed Monte Carlo (MC) simulations of the instrument in development. 
More specifically, CORSIKA \cite{CORSIKA-1998} is used for the simulation of the particle showers in the atmosphere and sim\_telarray \cite{Bernloehr-2008} for the detector simulation. 
The  candidates sites considered  are listed together with their altitude and geomagnetic field strength in Table \ref{table:sites}.
Besides these parameters, atmospheric profiles
derived specifically for each site are used in the air shower simulations.
The assumed night-sky background level (NSB)  corresponds to dark-sky observations towards an extra-galactic field and is set to identical values at all sites\footnote{
It should be noted that measurements by the CTA site work package revealed that the level of anthropogenic night sky background varies significantly between site candidates.}.
For a dedicated study, a small set of simulations has been carried out with elevated NSB levels (by 30\% and 50\%).

\begin{table}[htdp]

\begin{center}
\begin{tabular}{l|c|c|c|c}
Candidate site name & Latitude, Longitude & Altitude & B$_{\mathrm{hor}}$ & B$_{\mathrm{z}}\downarrow$ \\
       &                                &  [m]   & [$\mu$T] & [$\mu$T]\\
\hline\hline
Aar (Namibia) & $26.69^{\circ}$ S $6.44^{\circ}$ E & 1640 & 10.9 & -24.9 \\ 
Armazones (Chile) & $24.58^{\circ}$ S $70.24^{\circ}$ W & 2100 & 21.4 & -8.9 \\ 
Leoncito (Argentina)              & $31.72^{\circ}$ S $69.27^{\circ}$ W & 2640 & 19.9  & -12.6 \\
Leoncito++ (Argentina)     & $31.41^{\circ}$ S $69.49^{\circ}$ W& 1650 &19.9 & -12.6 \\
San Antonio de los Cobres  (SAC; Argentina)       &  $24.05^{\circ}$ S $66.24^{\circ}$ W   & 3600 & 20.9 & -8.9  \\
\hline
Meteor Crater (USA)  & $35.04^{\circ}$ N $111.03^{\circ}$ W & 1680 & 23.6 &  42.7 \\
San Pedro Martir (SPM; Mexico)  & $31.01^{\circ}$ N $115.48^{\circ}$ W & 2400 & 25.3 & 38.4 \\
Teide, Tenerife (Spain)  & $28.28^{\circ}$ N $16.54^{\circ}$ W & 2290 & 30.8 & 23.2 \\
\hline
Aar@500 m (hypothetical site) & $26.69^{\circ}$ S $6.44^{\circ}$ E & 500 & 10.9 & -24.9 \\
\end{tabular}
\end{center}
\caption{\label{table:sites} Summary table of information on the CTA candidates sites simulated.
The strength of the geomagnetic field is given by its horizontal (B$_{\mathrm{hor}}$) 
and downwards pointing (B$_{\mathrm{z}}\downarrow$) component.
Notes: 
i. Leoncito++: is a site candidate nearby the Leoncito site candidate in Argentina at an altitude of 1650 m.
ii. Aar@500m is a hypothetical site located at Aar, Namibia with an assumed altitude of 500 m. 
iii. The two site candidates located in the US (Yavapai and Meteor Crater) are very nearby each other and have similar characteristics. 
Therefore, only one of them has been simulated.
}
\end{table}

%

\begin{figure}[htbp]
\begin{center}
\centering\includegraphics[width=0.40\linewidth]{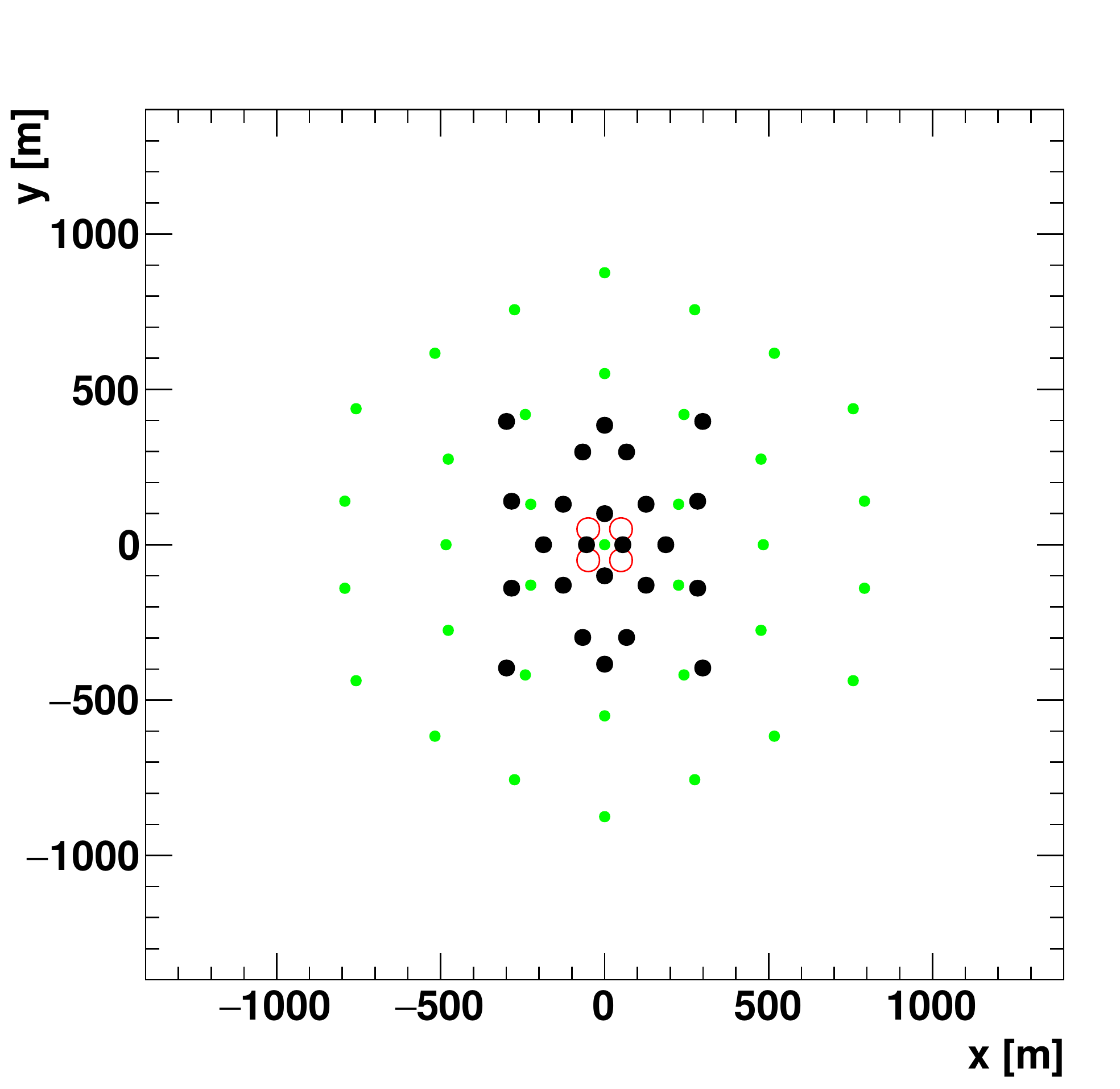}
\centering\includegraphics[width=0.40\linewidth]{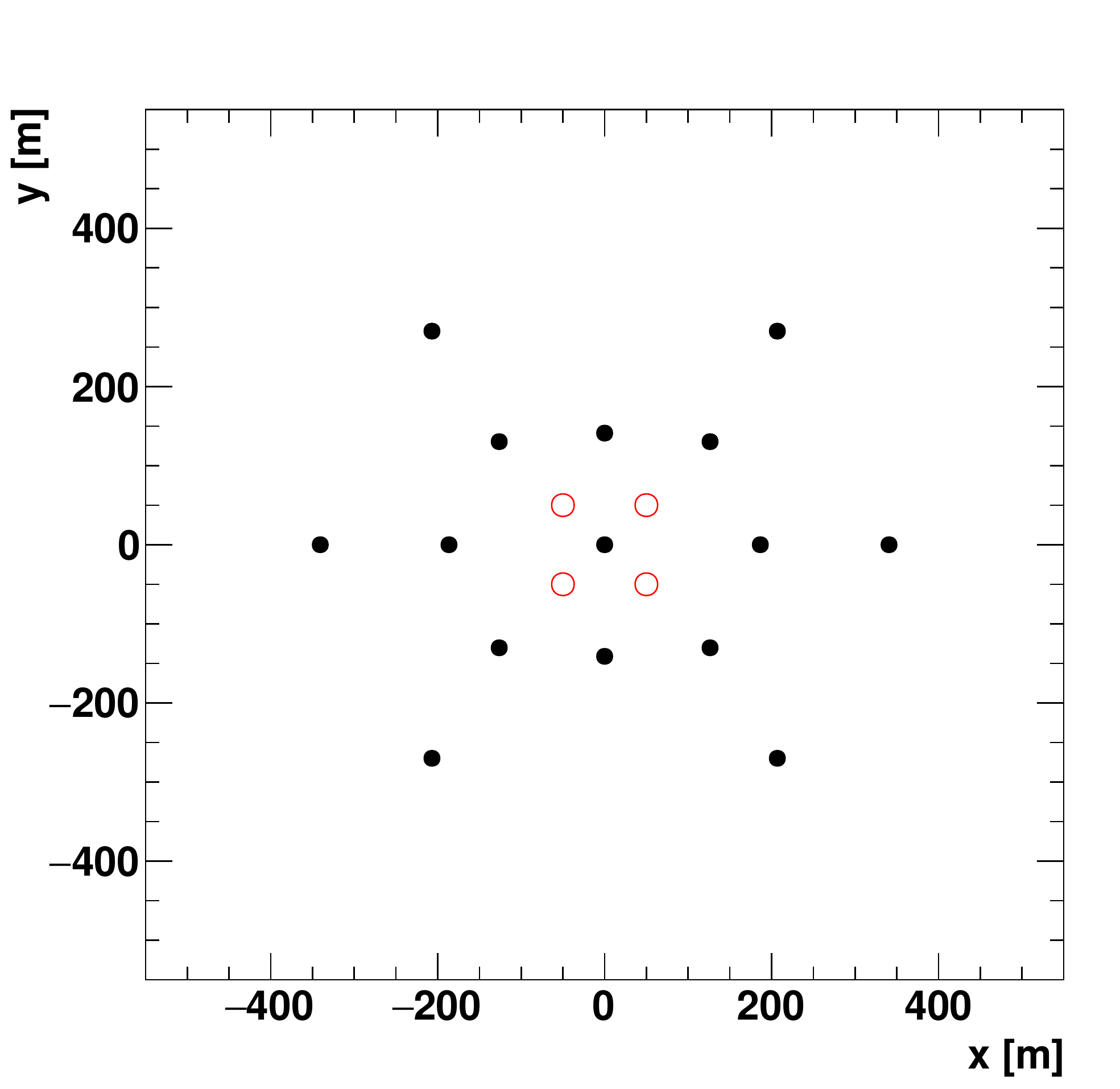}
\caption{Considered array layouts for CTA South (left; named S.2a) and CTA North (right; named N.2NN). 
Note the different scales of the axes between South and North.
Red circles: large-sized telescopes;
Black markers: mid-sized telescopes;
 Blue markers: small-sized telescopes.}
\label{Fig:Arrays}
\end{center}
\end{figure}

The details of the MC simulations, termed {\em prod-2}, are described in a separate contribution to this conference \cite{Hassan-2015}.
The telescopes, including cameras, trigger and readout electronics are assumed to be identical at all sites.
For each southern site, telescopes were simulated at a total of 229  positions, with many of the positions used for more than one telescope type. 
For some sites, the simulations include seven different telescope  types (LST, two MST types, and up to four SST types) while others include only four types (LST, MST plus two SST types, with SSTs not on as many positions as for the other sites). 
Unfortunately, the currently preferred layout with 4-m class SSTs depends on the additional SST positions and is thus not available for a comparison of all sites.
The array layout considered here consists of 4 LSTs, 24 MSTs, and 35 7m-class SSTs (see Figure \ref{Fig:Arrays}, left) for the southern sites, and of 4 LSTs and 15 MSTs (see Figure  \ref{Fig:Arrays}, right) for the layout for the northern sites. 
The conclusions on the science performance of the different sites have been tested to be independent of the chosen array layout.

The MC simulations consist of primary gamma-ray, proton, and electron events.
All results are for point-like gamma-ray sources located at the centre of the field of view and observed at a zenith angle of $20^{\circ}$ and 
two different azimuth directions (all telescopes pointing towards north or towards south).
A typical MC set for one site comprises about a billion simulated gamma-ray and electron, and about 100 billion proton events.
The simulation requires  substantial computing resources: the simulation of a single CTA site requires between 10-20 million HS06 CPU hours and 20 TBytes of event data written to disk.
A large fraction of the MC production used the EGI GRID, utilising the DIRAC framework as interware \cite{Arrabito-2014}.

\section{Analysis}

Several independent analysis packages derived from tools used in the H.E.S.S., MAGIC and VERITAS collaborations were used to process the MC production, see \cite{Bernloehr-2013,Hassan-2015} for details.
The results from the different packages are completely consistent with each other.
Here we present results obtained with the {\em eventdisplay} package\footnote{
One of the official analysis packages of the VERITAS collaboration, for details see {https://znwiki3.ifh.de/CTA/Eventdisplay\%20Software}}.

Briefly, this analysis consists of a two-pass FADC trace integration step \cite{Holder-2006}
using both charge and timing information to find the optimal placement of the trace integration window.
Pixels are selected using a two-level cleaning filter, searching for integrated signals
larger than five times their level of pedestal standard variation, or greater than four times the pedestal variation
for pixels adjacent to a pixel selected at higher threshold.
The resulting shower images are then parameterised by a second-moment ('Hillas') analysis, followed 
by a classical stereo reconstruction using the major  axes of the cleaned images.
The energy of each event is reconstructed using lookup tables of  mean gamma-ray energies
as a function of image amplitude (the sum of the integrated charges of the pixels selected in the image cleaning step)
and the distance of the telescope from the shower axis.
The gamma-hadron separation makes use of a boosted-decision tree algorithm, trained with a subset of the simulated signal and background events. 

CTA will be much more powerful in separating gamma rays from background hadronic events than current  observatories, as most of the events observed are sampled inside the area of the array.
These so-called {\em contained events} are much better sampled, providing improved background rejection, better angular resolution, and reduced energy threshold.
It is expected that the performance of the future CTA reconstruction pipeline will provide a significant improvement compared to the results presented in the following, as they are obtained with analyses optimised for current instruments with 2-5 telescopes in operation.

\section{Science Performance}


\begin{figure}[htbp]
\begin{center}
\centering\includegraphics[width=0.99\linewidth]{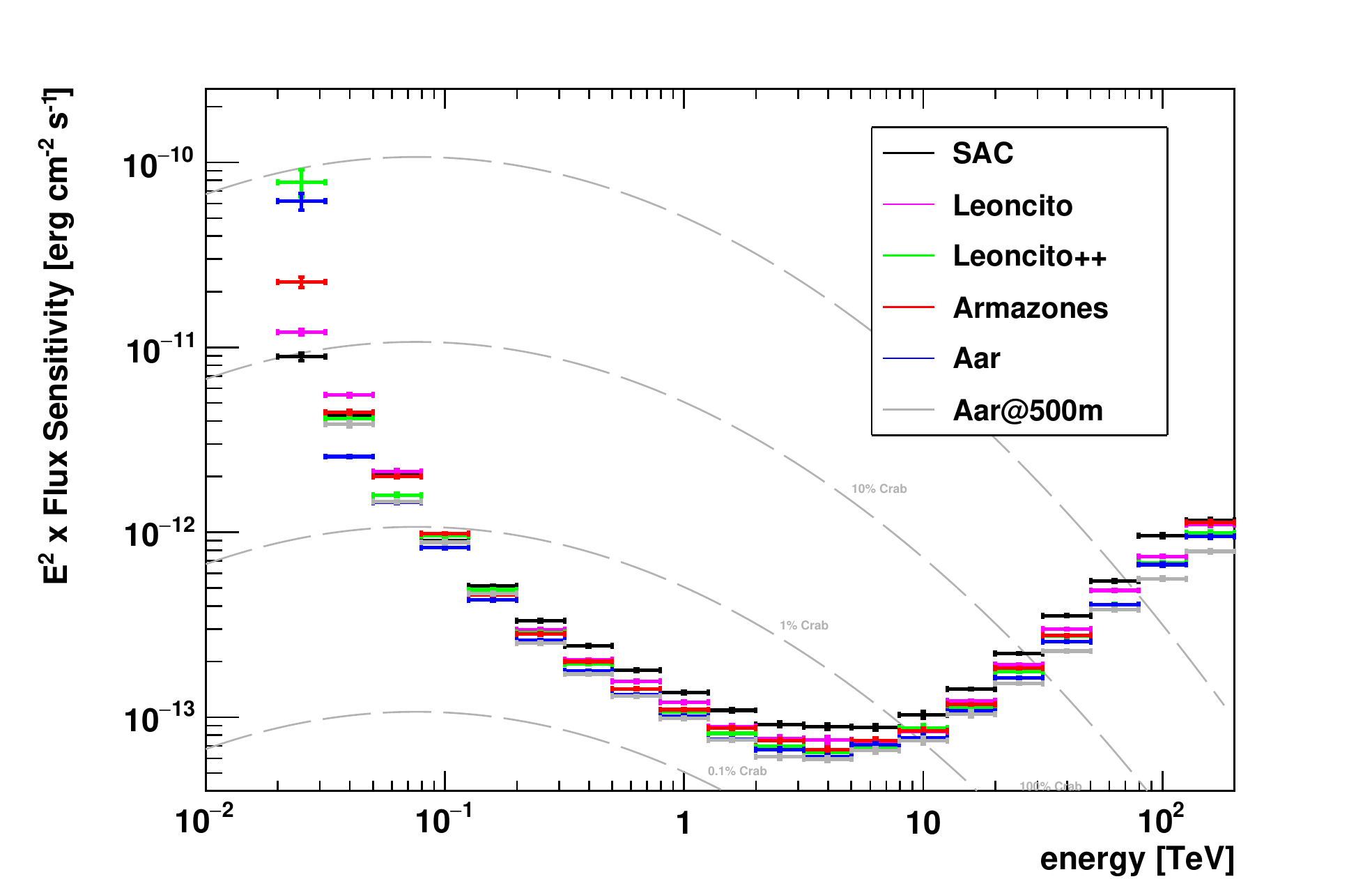}
\centering\includegraphics[width=0.99\linewidth]{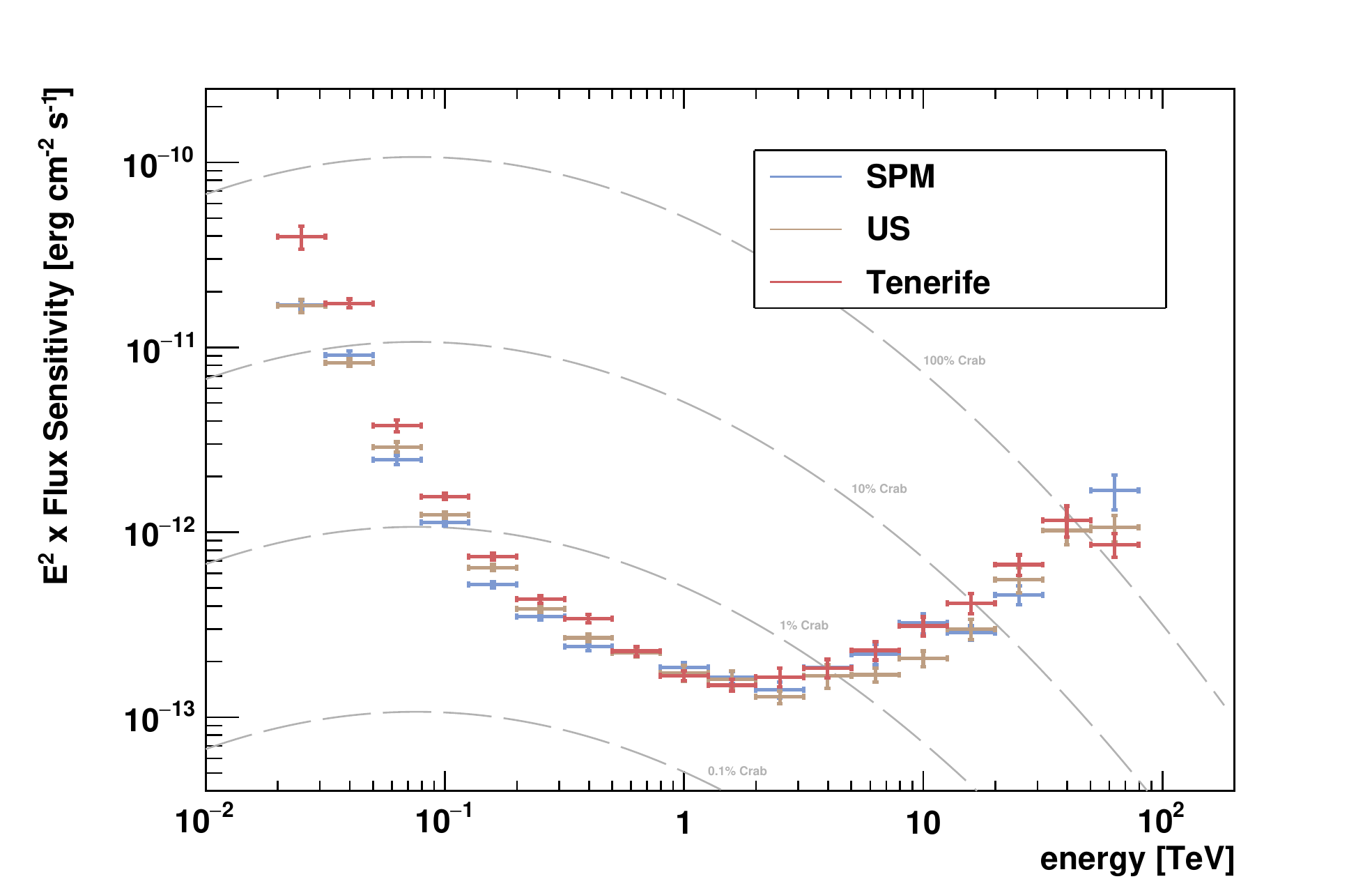}
\caption{On-axis differential point-source sensitivity for the considered CTA site candidates located in the Southern (top)
and Northern hemisphere (bottom)
(see Table 1). 
Average sensitivities calculated from telescopes pointing towards north and south are shown.
The layout candidate 'S.2a' has been used for CTA South, 'N.2NN' for CTA North.
Differential sensitivities are derived for 50 h of observations.
The dashed lines indicate the flux of a Crab Nebula-like source scaled by the factors indicated in the figure,
horizontal 'error' bars indicate the bin size in energy.
}
\label{Fig:DiffSens}
\end{center}
\end{figure}

%

\begin{figure}[htbp]
\begin{center}
\centering\includegraphics[width=0.40\linewidth]{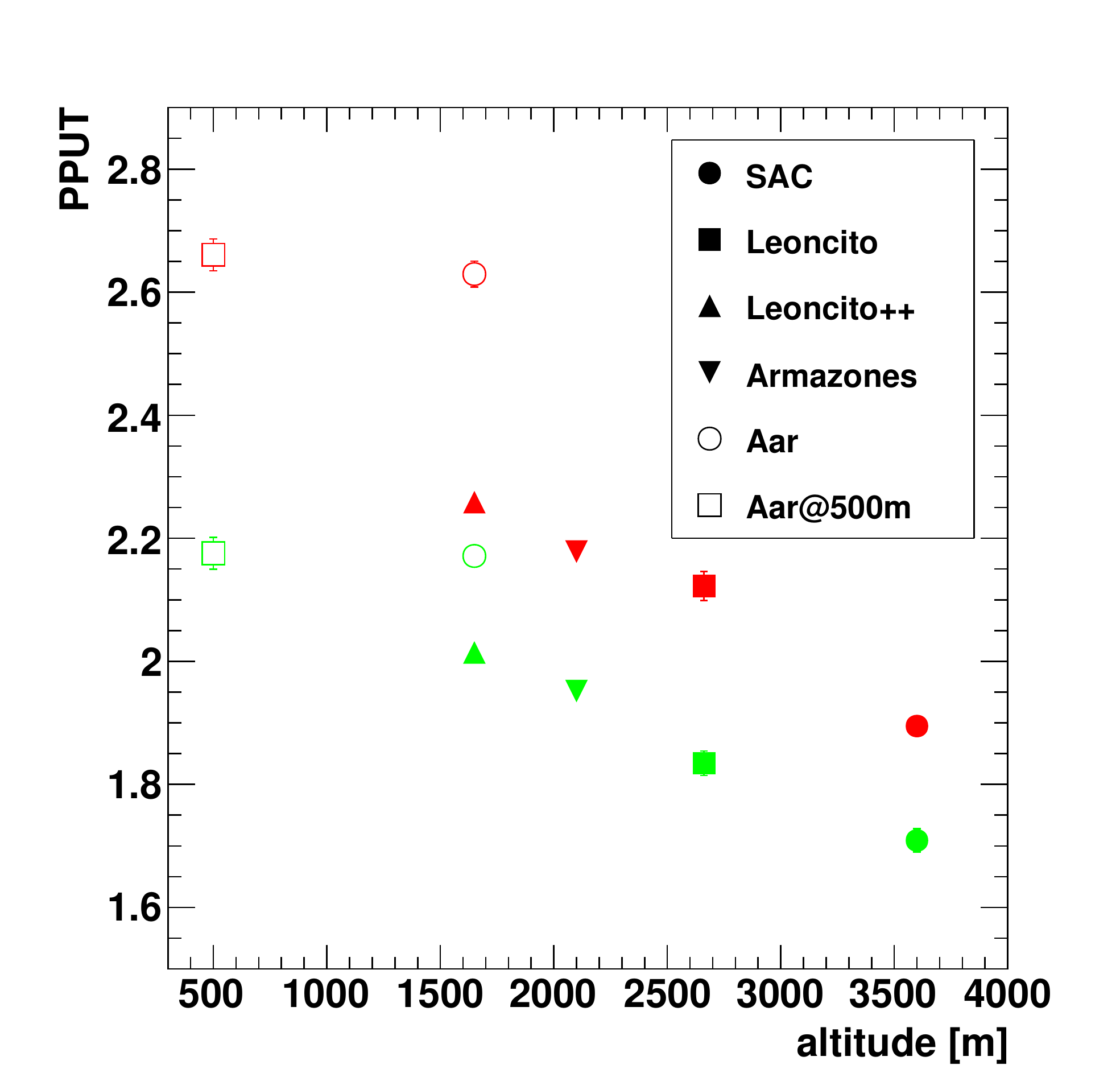}
\centering\includegraphics[width=0.40\linewidth]{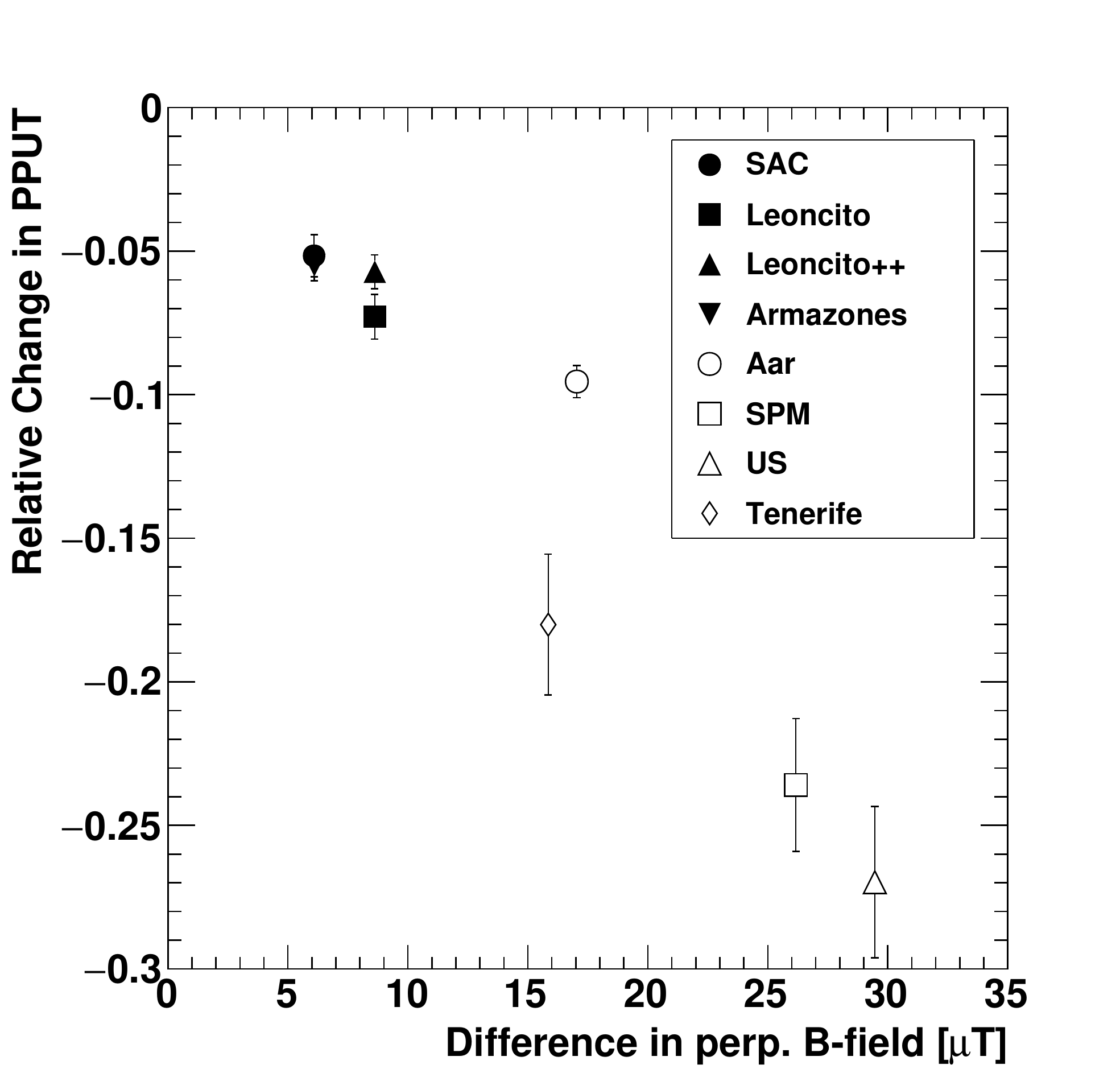}
\caption{Left: Performance per unit time (PPUT; see text for definition) as function of altitude for the site candidates located in the Southern hemisphere.
Red-colored markers indicate results for the arrays pointing in southerly directions, green markers pointings towards the North.
Right:  Relative difference in PPUT (defined as (PPUT1-PPUT2)/(0.5*(PPUT1+PPUT2))) for pairs of simulations at the same site, pointing at $20^{\circ}$ zenith angle to the North and the South, versus difference in perpendicular B-field strength between these two directions. 
Northern and southern site candidates are included in this figure. 
The differences in the size of the uncertainties is due to the much lower number of MC events simulated for the northern site candidates.}
\label{Fig:PPUT}
\end{center}
\end{figure}

The primary performance criteria for the site evaluation are differential sensitivity over the entire energy range of CTA from 30 GeV to 300 TeV.
The differential sensitivity, i.e.~the sensitivity to point-like gamma-ray sources calculated for a small range of energies,  incorporates collection area, angular resolution, and background rate.
It is defined as an independent detection (5 sigma significance, $\ge$10 excess events, and more than 5\% of the remaining background) in each energy bin. We use five logarithmic bins per decade of energy.
 
As good sensitivity is required over the complete energy range defined above, the figure of merit used for easier comparison of the science performance at the different site candidates is the so-called performance per unit time (PPUT).
It is defined as the inverse of the geometric mean of the sensitivity in individual energy bins, normalised by a reference sensitivity in each band:
\begin{equation}
\mathrm{PPUT} =  \left( \prod_{N bins} \frac{F_{sens, ref}}{F_{sens}} \right)^{1/N}
\end{equation}
with $N$ bins in energy (from 30 GeV to 300 (20) TeV for CTA South (North)) of the reference sensitivity $F_{sens, ref}$ and the achieved flux sensitivity $F_{sens}$.
The reference sensitivity $F_{sens, ref}$ has been derived from the analysis of a previous production of CTA simulations ({\em prod1}, see \cite{Bernloehr-2013}) for a site at 2000 m altitude and with a geomagnetic field strength and orientation intermediate between that found at the Aar and Tenerife sites.
This first production of MC simulations for CTA was based on initial and conservative assumptions of telescope parameters and without the possibility of pursuing an advanced analysis of the FADC traces, which improves especially the sensitivity to low-energy events.
For these reasons, PPUT values are expected to be significantly larger than one.
 
 \subsection{Performance for dark-sky observations}
 
\begin{figure}[htbp]
\begin{center}
\centering\includegraphics[width=0.40\linewidth]{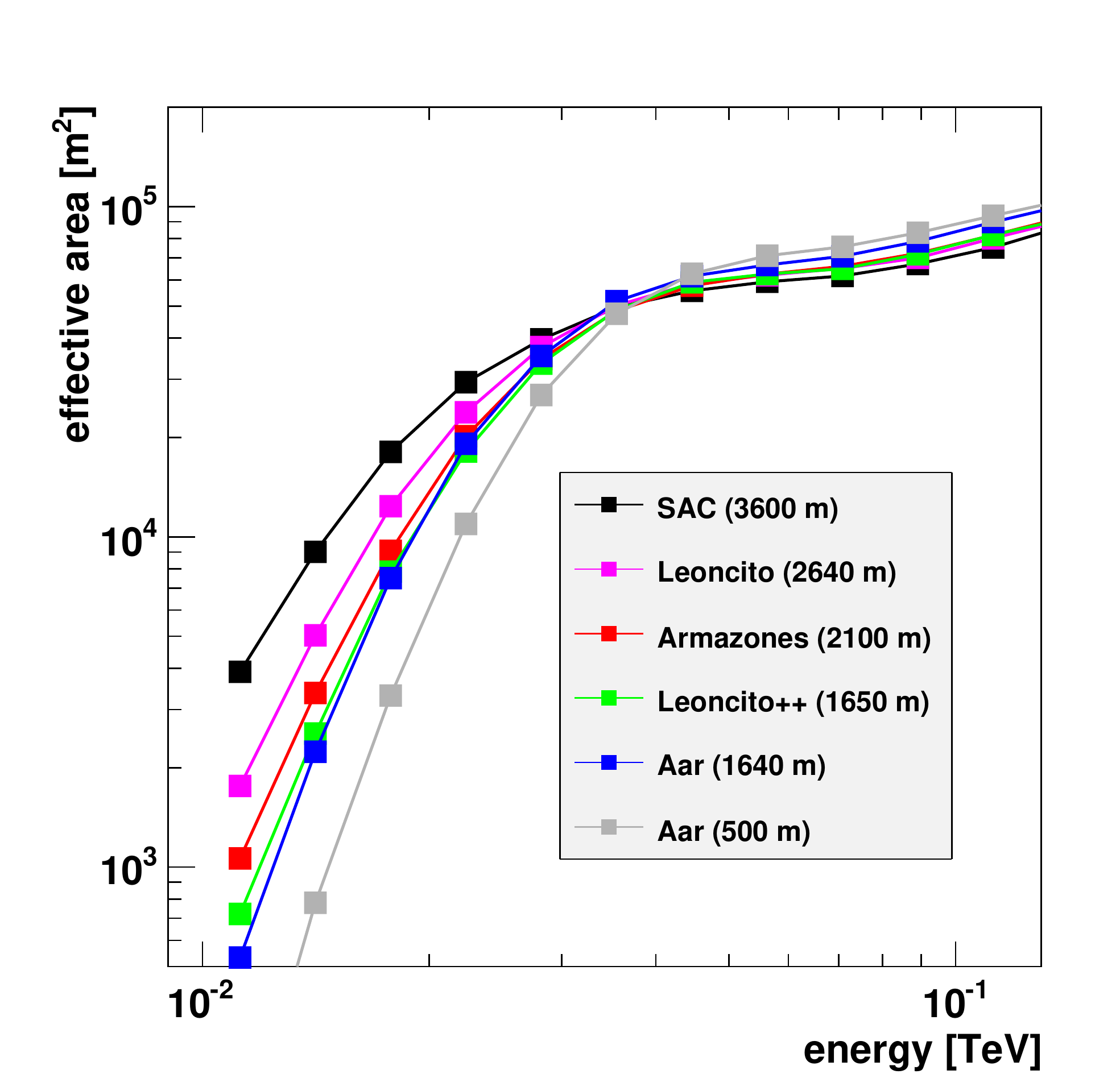}
\centering\includegraphics[width=0.40\linewidth]{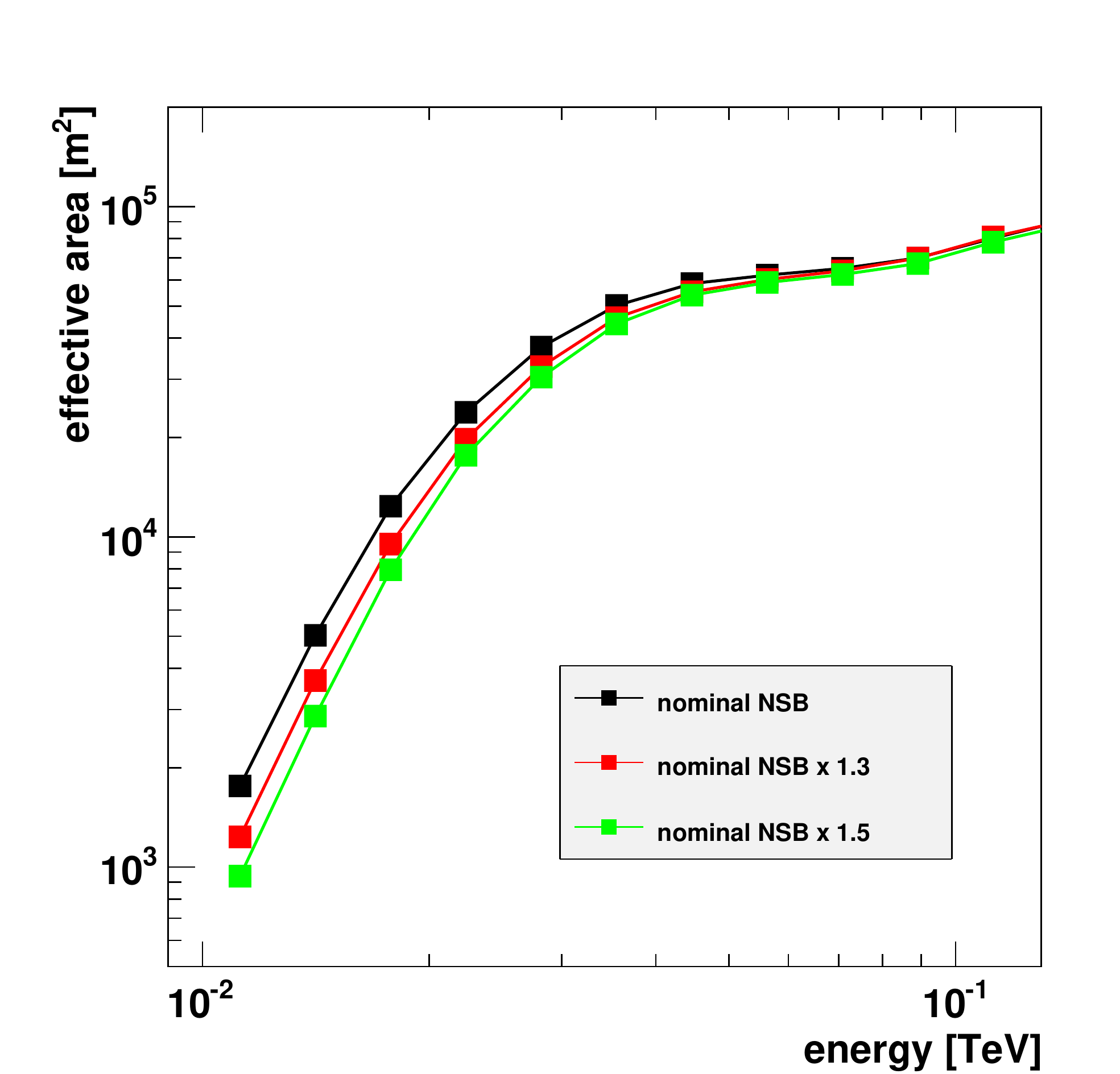}
\caption{Left: Effective areas vs true energy in the threshold region for the site candidates located in the Southern hemisphere. 
Right: Effective areas vs true energy in the threshold region for different levels of night-sky background light for the Leoncito site.
The effective area of the instrument is the differential gamma-ray detection rate dN/dE after cuts, divided by the differential flux of incident gammas.
Very loose cuts have been applied in the analysis for both figures, requiring the successful reconstruction of direction and energy only.
}
\label{Fig:EffAreas}
\end{center}
\end{figure}

Figure \ref{Fig:DiffSens} shows the on-axis differential point-source sensitivity vs energy for the considered CTA site candidates,
Figures \ref{Fig:PPUT} and \ref{Fig:EffAreas} the effect of site altitude and geomagnetic field on the overall sensitivity and the response at threshold energies. 
Clearly, science performance varies significantly between the sites.
At the {\bf energy threshold} of the instrument, the detection is limited by the number of Cherenkov photons hitting the telescopes.
Higher-altitude sites like the San Antonio de los Cobres site at 3600 m a.s.l.~show the best performance among all sites at energies in the 20-40 GeV range; they are closer to the shower maximum with higher photon intensities on the ground at distances $<$150 m to the shower axis 
(see especially Figure \ref{Fig:EffAreas}, left).

The mid-energy range (roughly from 50 GeV to 10 TeV for CTA South respectively 5 TeV for CTA North) is {\bf background dominated} and performance is limited by the ability to suppress background events and  by the quality of the angular reconstruction.
Altitude affects sensitivity in this energy range in two ways: for a given gamma-ray energy, the predominantly proton background appears more gamma-ray like at higher altitude; and very close to the shower axis the contribution from particles penetrating to ground level increases with altitude, increasing the level of fluctuations in gamma-ray images and complicating the gamma-hadron separation.
Lower-altitude sites clearly perform better in this energy range. 
This was the reason to explore in addition to the CTA candidate sites also a hypothetical site at 500 m altitude. 
The Aar@500m site shows  good performance in the mid-energy range, but not better than the Aar site at 1640 m altitude.
The benefits of lower-altitude sites described above seem to be balanced by a decreased number of photons, mainly due to the larger distance of the observatory to the shower maximum. 
The significantly lower effective area at energies below 50 GeV of the (hypothetical) Aar@500m site would deteriorate particularly its ability to detect soft transients like Gamma-ray bursts.

The geomagnetic field produces a separation of positive and negative charges in air showers, leading to a distortion of the Cherenkov light pool on the ground and of the shapes of images.
This affects gamma-hadron separation, the angular and energy reconstruction.
Figure \ref{Fig:PPUT} reveals significant differences in performance between observing showers with directions close to parallel to the geomagnetic field lines (arrays pointing south for sites in the Southern hemisphere) to observations in directions at larger angles to the  field lines.
The sensitivity of the sites in South America is inferior to those in Southern Africa, at equal altitude, mainly because the different inclination of the  geomagnetic field results in larger average field strengths perpendicular to the shower direction at small zenith angles in South America. 
For larger zenith angles (simulations being available for some sites at 40$^\circ$) the differences are smaller.
It should be noted that the considered zenith angle of $20^{\circ}$ is the worst case scenario for the Northern sites, resulting in showers propagating almost exactly perpendicular to the geomagnetic field lines.
The average zenith angle for observations for CTA will very likely be around $30^{\circ}$.
At the highest energies, the sensitivity is {\bf limited by the signal collection area}, which is larger at low-altitude sites due to the higher density of Cherenkov photons at large impact distances.

\subsection{Performance at increased night-sky background levels}

Higher night-sky background levels are expected at some sites due to increased anthropogenic night levels, but also at all sites
for observations towards bright regions in the sky, e.g.~toward the Galactic Centre region.
The level of night-sky background (NSB) light affects the performance mainly in the threshold region.
Higher NSB produces higher accidental rates, requiring increased trigger thresholds.
It also lowers the signal-to-noise ratio in images, leading to a loss of low-energy events and quality in reconstruction.
Figure \ref{Fig:EffAreas}, right, shows the impact of increasing the NSB level by 30\% and 50\% on the effective areas for the Leoncito site. 
At 40 (25) GeV, effective areas are between 10 and 20 \% (50 and 70\%) lower for higher NSB levels.

\section{Conclusions}

The science performance of the CTA candidate sites differ significantly, showing a strong dependence on site altitude and strength of the geomagnetic field.
The best overall performance is expected for sites at around 1700 m site altitude, with an acceptable range of altitudes between 1600 and 2500 m.
In the meantime, the CTA site selection process is progressing rapidly: Government representatives of the countries involved in CTA shortlisted two sites for each hemisphere based on the input from the CTA collaboration on annual available observing time, science performance (this study), risks, and cost. 
The two sites in the Southern Hemisphere are Aar and Armazones (with Leoncito in Argentina kept as a third option), for the Northern Hemisphere the selected sites are Tenerife and San Pedro Martir.


\subsubsection*{Acknowledgments}
We gratefully acknowledge support from the agencies and organizations 
listed under Funding Agencies at this website: http://www.cta-observatory.org/.


\end{document}